# Rocketquake Seismology with a Falcon 9 Rocket Source

G.T. Schuster (University of Utah), J. Li (Dept. of Geophysics, Jilin University), S. Hanafy (King Fahd University of Petroleum and Mining), V. G. Thompson (Carbon Energy Corporation), J. Farrell (University of Utah), and D. Trentman (University of Utah)

*Email: inter_lijing@jlu.edu.cn; gerard.schuster2022@gmail.com*

## Abstract

We investigate the feasibility of using rocket launches, specifically rocketquakes, as a seismic source to image subsurface velocity and geology of planetary bodies. Toward this goal, we record the seismic vibrations excited by a Falcon 9 rocket launch from Vandenberg Space Force Base (SFB) near Lompoc, California. Nine passive three-component (3C) seismometers were deployed every 18.75 meters along a 45-degree line from the launch site starting at the offset of about 7 km kilometers from the launch pad. Results show that coherent body waves can be recorded with a P-velocity of more than 2.0 km/s and a penetration depth of 1 km or deeper. Stronger Rayleigh waves were also recorded and inverted to give an S-velocity profile to a depth of 60 m. The imaging techniques employed for rocketquake seismology integrate inversion methods from earthquake and exploration seismology, yielding the P- and S-velocity profiles of the subsurface. These results suggest that rocket launches as seismic sources will provide unprecedented opportunities for identifying the subsurface hazards, faults, tunnels, water ice, and mineral deposits of planetary bodies and their moons.

## Introduction

According to a synopsis in the Financial Times (FT) on October 18, 2023, there is a growing race to establish lunar bases, create a lunar economy, and commercialize the Moon's resources, with major players including the USA, China, Russia, India, and Japan. These resources range from minerals and tourism to scientific observatories and strategic interests. NASA has planned the lunar orbiter Gateway (https://www.nasa.gov/mission/gateway/), which will serve as a hub for astronauts visiting lunar bases. The FT article estimates that over the next decade, more than $136 billion will be invested in developing a lunar economy. Additionally, over $100 billion is expected to be spent on lunar transportation and logistics by 2040. The successful launch of a Starship on October 13, 2024 (https://www.nytimes.com/live/2024/10/13/science/spacex-starship-launch), signals the possibility of a manned mission to Mars within the next 15 years and the establishment of rocket bases on the Red Planet.

A critical resource for establishing a viable lunar or Martian economy is the discovery and extraction of water from subsurface deposits. The European Space Agency recently announced the potential discovery of water-ice deposits on Mars tons (https://esa.int/Science_Exploration/Space_Science/Mars_Express/Buried_water_ice_at_ Mars_s_equator). Similarly, water ice is believed to exist at the lunar South Pole; estimates suggest as much as 6 billion (https://www.weforum.org/stories/2023/08/space-water-ice-moon-south-pole/#:). This water can be utilized to extract oxygen and hydrogen-oxygen for life support systems and hydrogen as an energy source for rockets, which could eventually be used for missions to Mars or asteroids. Advanced geophysical imaging methods are essential to locate these valuable water and mineral resources. On Earth, seismic and resistivity imaging are the primary techniques for detecting water in the subsurface. Ground-penetrating radar can also be effective for water detection if salty minerals are absent.

Seismic imaging is widely regarded as one of the most effective methods for detecting unstable geology, faults, and fluid resources. Identifying unstable subsurface conditions is essential to ensure the safety of planetary bases and planned nuclear reactors. However, one challenge of seismic imaging is the high cost of transporting large seismic sources to the Moon. Is there a more cost-effective alternative in the early stages of lunar exploration? Yes, rockets such as the Falcon 9 (https://en.wikipedia.org/wiki/Falcon_9) shown in Figure 1, could serve as a viable solution!

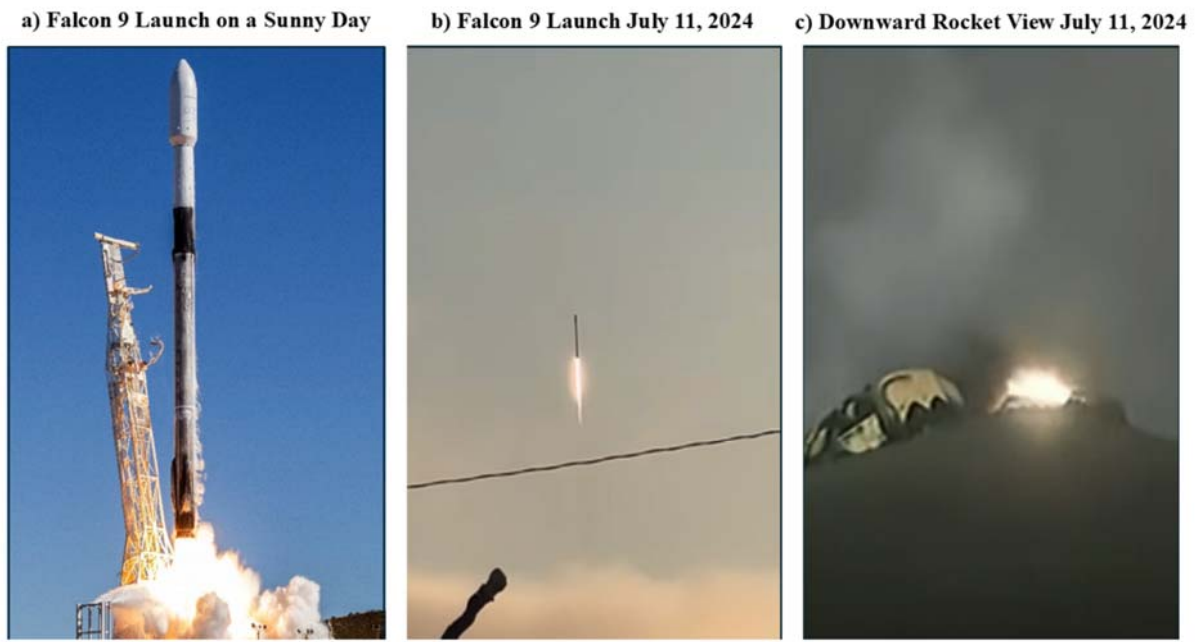

**Figure 1:** Launch of a Falcon 9 rocket from Vandenberg SFB, California a) on a sunny day, b) on July 11, 2024, and c) viewed from the tail of the rocket. The a) image is from Matthews et al. (2020), the b) image is from the town of Lompoc July 11, 2024 (David Bull, https://www.facebook.com/groups/233130583862628/posts/1831931600649177/), and c) is the image from the SpaceX camera mounted on the tail of the rocket as it launched July 11 through the fog.

**Goal**

The goal is to evaluate the feasibility of using seismic vibrations generated by a Falcon 9 launch at Vandenberg Space Force Base (SFB) to estimate subsurface velocity structures. Previous studies[1,2] have shown that the launch or landing of a Falcon 9 rocket produces strong acoustic airwaves, ranging from nearly 0 Hz to over 100 Hz, which can be detected over distances greater than 10 km from the rocket. Additionally, the exhaust gases from the rocket can generate significant subsurface vibrations by impacting the ground near the launch pad. In our study, the spectral signature of the infrasound recordings from Mathews[1] and Durant[2] closely resembles the bandwidth of seismic recordings. These findings suggest that, for future missions to Mars or the Moon, similar seismic data could be used to analyze lunar and Martian geology and assess the presence of subsurface water, lava tubes, ice, mineral deposits, and potential geologic hazards.

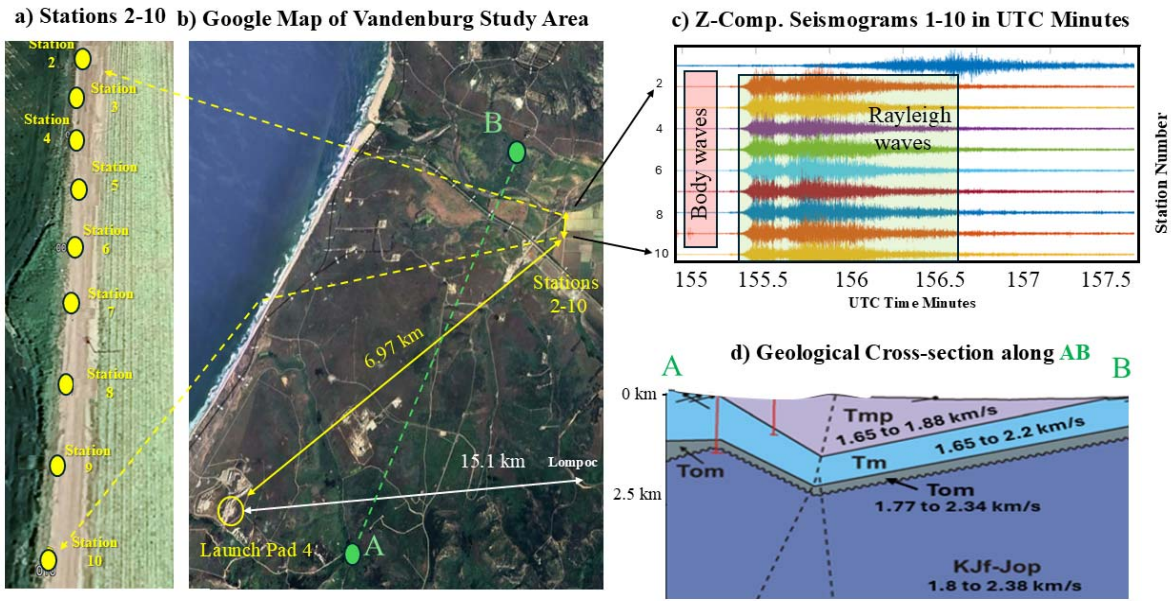

**Figure 2. Google map, recorded data, and geologic cross-section. a) Locations of geophones 2-10 at 18.75 m intervals and b) Google map of launch area at VSFB, c) Z-component seismograms recorded at stations 1-10, d) geological cross-section along AB.**

Schuster deployed and recorded the seismic vibrations from a Falcon 9 rocket launched Thursday, July 11 at 7:35 PM local California time (0235 UTC) at the Vandenberg SFB in California. Nine of the three-component (3C) recorders in Figure 2 are about 7 km from the launch site while the other recorder was located about 15 km away in Lompoc. The recorders are ten 3C Fairfield recorders. The infrasound recording site of Mathews[1] is about 1 km to the east of stations 2-10. where it recorded the sound pressure in the air for 3 different Falcon 9 launches at different times of the year and different weather conditions; the infrasound recorder was not active for the July 11, 2024, launch.

## Results: Seismic Data and Interpretation

The recorded Z-component seismograms are shown in Figure 2c, where the large amplitude arrivals from 155.4-156.5 minutes (green transparent window) for traces 2-10 are mainly due to the Rayleigh waves generated from the rocket launched at 155 minutes UTC (7:35 PM California time). Using an acoustic propagation velocity of 0.340 km/s[1], the time to travel 7 km from launch pad 4 to recorders 2-10 is 7/0.34 ≈ 20 s, which is 20/60 = 0.33 minutes from the rocket's ignition time at 155 minutes. As seen in Figure 2c, the visible train of arrivals starts to build up around this time. Coincidentally, the shallow subsurface S-wave velocity varies between 150 to 350 m/s according to the inversion of the Rayleigh waves generated by the exhaust thrust of the rocket impacting the launch pad and, to a lesser extent, air-coupled Rayleigh waves[3]. The spectra for the Z-component recordings show useful information from about 0.2 Hz to more than 50 Hz (see Figure 1 in Appendix).

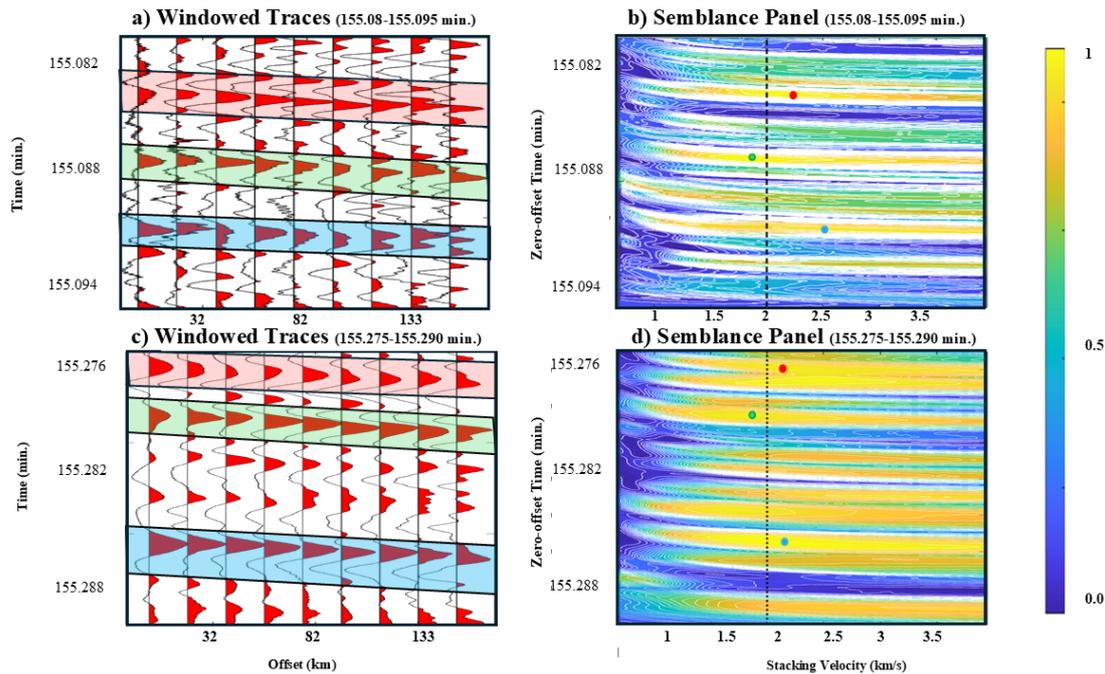

**Figure 3.** Seismograms and computed velocity semblance panels. a) and c) are seismograms with body waves recorded at stations 2-10. These seismograms are used to compute the velocity semblance panels on the right. The red, green, and blue dots in the semblance panels indicate the maximum semblance values of coherent events in associated seismograms.

**Crustal Body Waves**

Figures 3a and 3c depict the Z-component seismograms recorded within the red transparent rectangle in Figure 2c). The velocity semblance values[4] are computed from these seismograms, and the resulting semblance panels in Figures 3b and 3d indicate that the maximum semblance values have a moveout velocity of between 1.9-2.5 km/s. The red, green, and blue dots in the semblance panel of Figure 3b indicate the maximum semblance values of coherent events in Figure 3a. In this case, the original moveout velocity was corrected by the multiplicative factor of $\cos(\pi/4)$ because the recording line is slanted at approximately a 45-degree angle to the radial line from the launch pad. Coherent arrivals with similar moveout velocities appear at time intervals of about 0.002 minutes. These are reverberating arrivals excited by the pulsating nature of the rocket motor that produced exhaust gases and airwaves that impacted the ground. These reverberations produce replicated body-wave arrivals that can be used to check for consistency of interpretation and estimate the uncertainty in the propagation velocity. According to the geologic cross-section in Figure 2d[5,6], these events with a moveout velocity of around 2.1 km/s are crustal body waves that dive 1-2 km into the crust and return with about the same apparent velocity as the deepest layer

**Rayleigh Waves**

An ascending rocket generates reverberating exhaust gases and acoustic waves that strike the ground to excite propagating Rayleigh waves in the subsurface. Acoustic waves striking the ground excite Rayleigh waves[3,7], where the coupling between the airwave and excited Rayleigh wave is strongest when the airwave velocity is close to that of the Rayleigh wave velocity at the near-surface. These waves can be modeled as a weighted sum of Green's functions excited by a string of ascending point sources[8].

Experiments with firecracker explosions and five rocket explosions were analyzed by Novoselov[9] using collocated seismic and infrasound sensors; they found that around 2 percent of the acoustic energy is admitted into the ground (converted to seismic energy). They also found that the Rayleigh wave, which propagates more than twice as fast as the acoustic wave, precedes the infrasound airwave at far-offset receivers. This suggests that early arrivals recorded from a rocket source might be windowed to exclude the strong air-coupled Rayleigh waves and acoustic arrivals. For seismic recordings on Earth and Mars, the exhaust gases that strike the ground at the early portion of the launch will be the largest contributor to propagating body waves and Rayleigh waves. Unlike Earth or Mars, there are no acoustic waves on the moon so the exhaust gases impinging on a lunar launch pad will be the exclusive generator of seismic waves.

**Wave Equation Dispersion Inversion of Rayleigh Waves**

The fundamental dispersion curve associated with the Rayleigh waves can be inverted for the S-velocity[10]. Figure 4a displays the virtual seismograms obtained by 1) correlating the Z-component traces 2-10 within the recording window of 155 to 175 minutes. Trace #2 is correlated with traces 2-10, and the negative-lag-time correlograms in Figure 4a were flipped in polarity, mirrored across the zero-lag time (dashed black line in a) ), and added to the positive time events[10]. These stacked traces were windowed to exclude the acausal events, and then the phase velocity image in Figure 4b was computed by a high-resolution Radon transform (HRT). The maximum amplitudes in the phase velocity image were used to define the green dispersion curve in Figure 4b. This interferometry procedure produced a similar dispersion curve when applied to the Z-component seismograms recorded between 155 and 157 minutes.

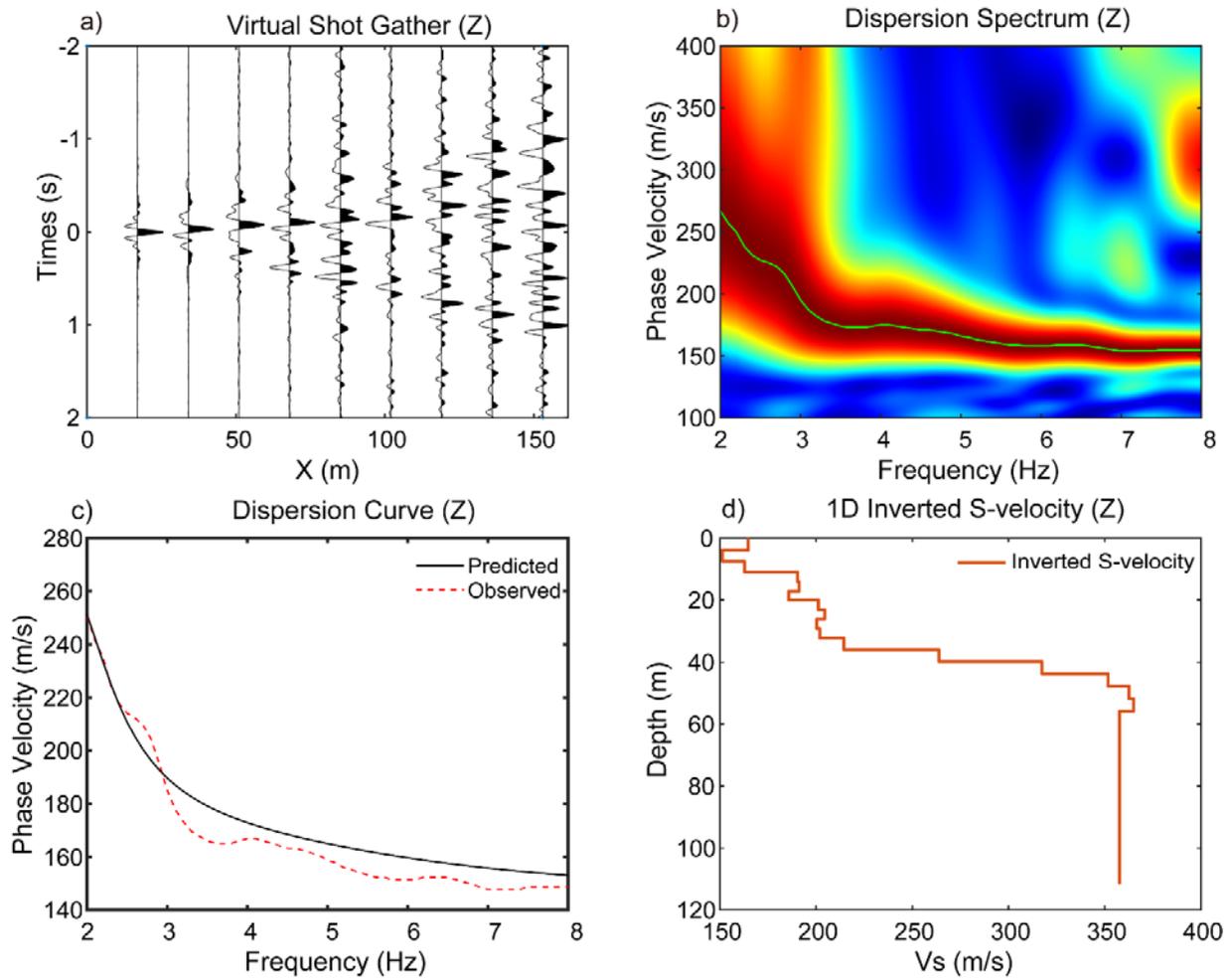

**Figure 4. Results of analyzing the Z-component seismograms. a) Virtual Rayleigh waves, the b) dispersion spectrum, c) predicted and observed dispersion curves, and d) inverted S-velocity model. All of these results are computed from the Z-component seismograms in Figure 2c.**

The reliability of the inversion profile was tested using the strategy of Yan[11]. The layered model is divided into 10-m thick layers, and the number of layers is adjusted adaptively according to the penetration depth of the dispersion curve (1/2 wavelength). We used the Dix-type inverted result as the initial velocity model for the preconditioned fast-descent inversion. Figure 4c compares the predicted (solid line) and observed (dashed line) dispersion curves, where the predicted dispersion curve is obtained from the inverted velocity model in Figure 4d. This final 1D S-velocity was obtained after 20 iterations. The long-wavelength trend of the observed (dashed line) and inverted (solid line) dispersion curves largely agree with one another at all frequency ranges.

As a consistency check, the NS-component seismograms were used to interferometrically compute the virtual NS seismograms, which were then used to compute the dispersion curve and S-wave velocity model. The results are shown in Figure 5, where the S-wave velocity model in Figure 5d is largely consistent with that in Figure 4d.

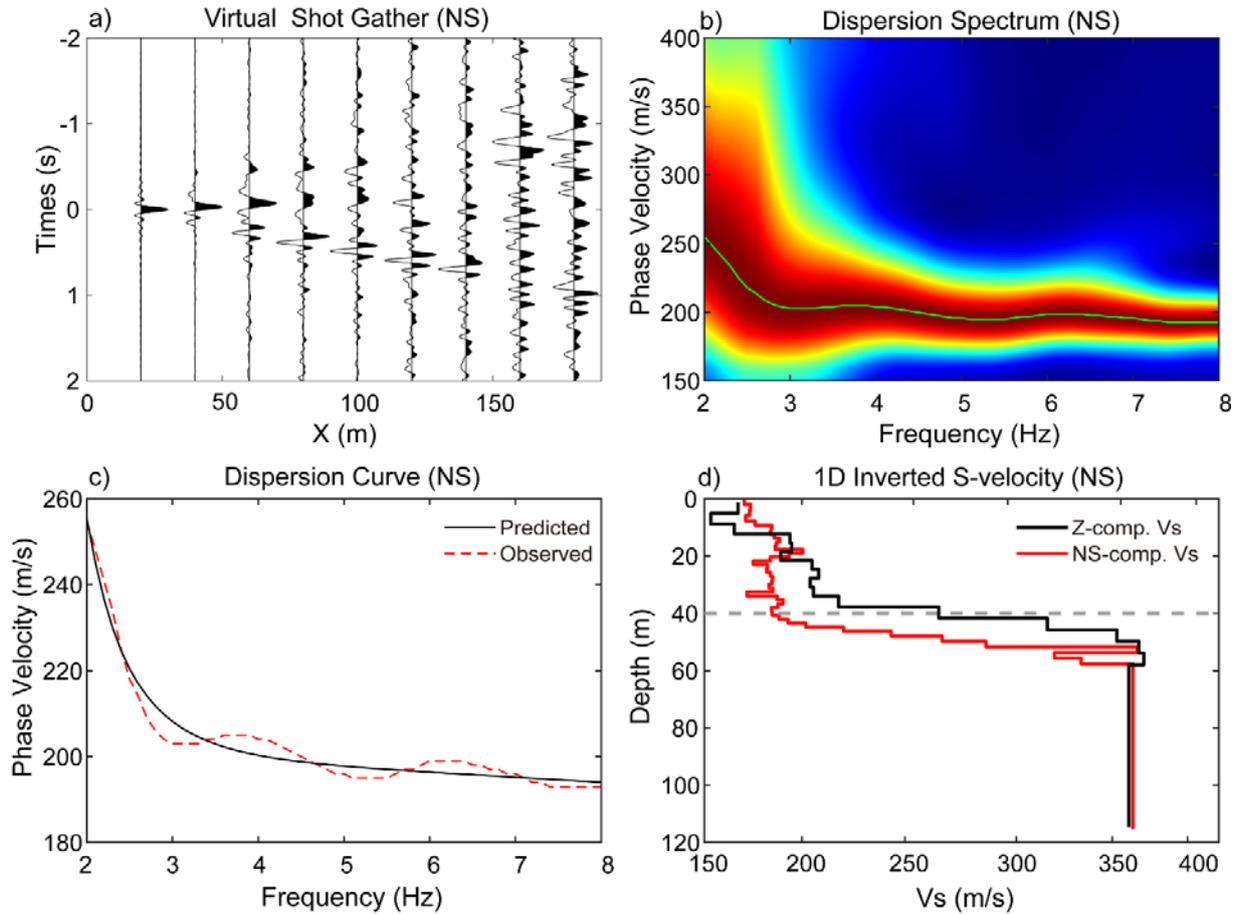

**Figure 5.** Results of analyzing the NS-component seismograms. a) Virtual Rayleigh waves along NS-component, b) dispersion spectrum, c) predicted and observed dispersion curves, and d) the inverted S-velocity profile denoted by the red curve. The solid black curve in d) is the same as the orange velocity profile in Figure 4d.

**P-velocity Profile from Pre-ignition Data**

Deploying the geophones before launch presents opportunities for recording seismograms generated by hammering the geophones into the ground. As an example, three common shot gathers (CSGs) were recorded by hammering on a geophone before the ignition time, with all the impacts located at receiver no. 10. Figures 6a-6b show two of these three common shot gathers as an example. The first arrival travel times (FATs) are distinctly visible in the recorded data. We manually picked the FATs for all three CSGs, as illustrated in Figure 6c, which shows a close agreement of the FATS from each CSG. The FATs exhibit two discernible slopes indicative of distinct subsurface layers.

Inspecting the recording stations by walking along the recording line can produce events that can be interferometrically processed to produce virtual shot gathers. Similar data can be created by driving a Moon or Mars buggy up and down along a recording line (see Figure 2 in Appendix). This suggests that any geophone deployment on the moon or Mars can use impulsive noise induced by the astronaut to determine the near-surface velocity and detect the presence of shallow faults or lava tunnels. A demonstration of this capability was performed by Hanafy[12] who continuously drove a vehicle along a line of passive recorders for about 30 minutes. They applied the interferometric procedure of correlation and stacking of traces within overlapping windows of the recorded data. The resulting virtual shot gathers revealed surface waves with moveout velocities that closely approximated those from active source shot gathers.

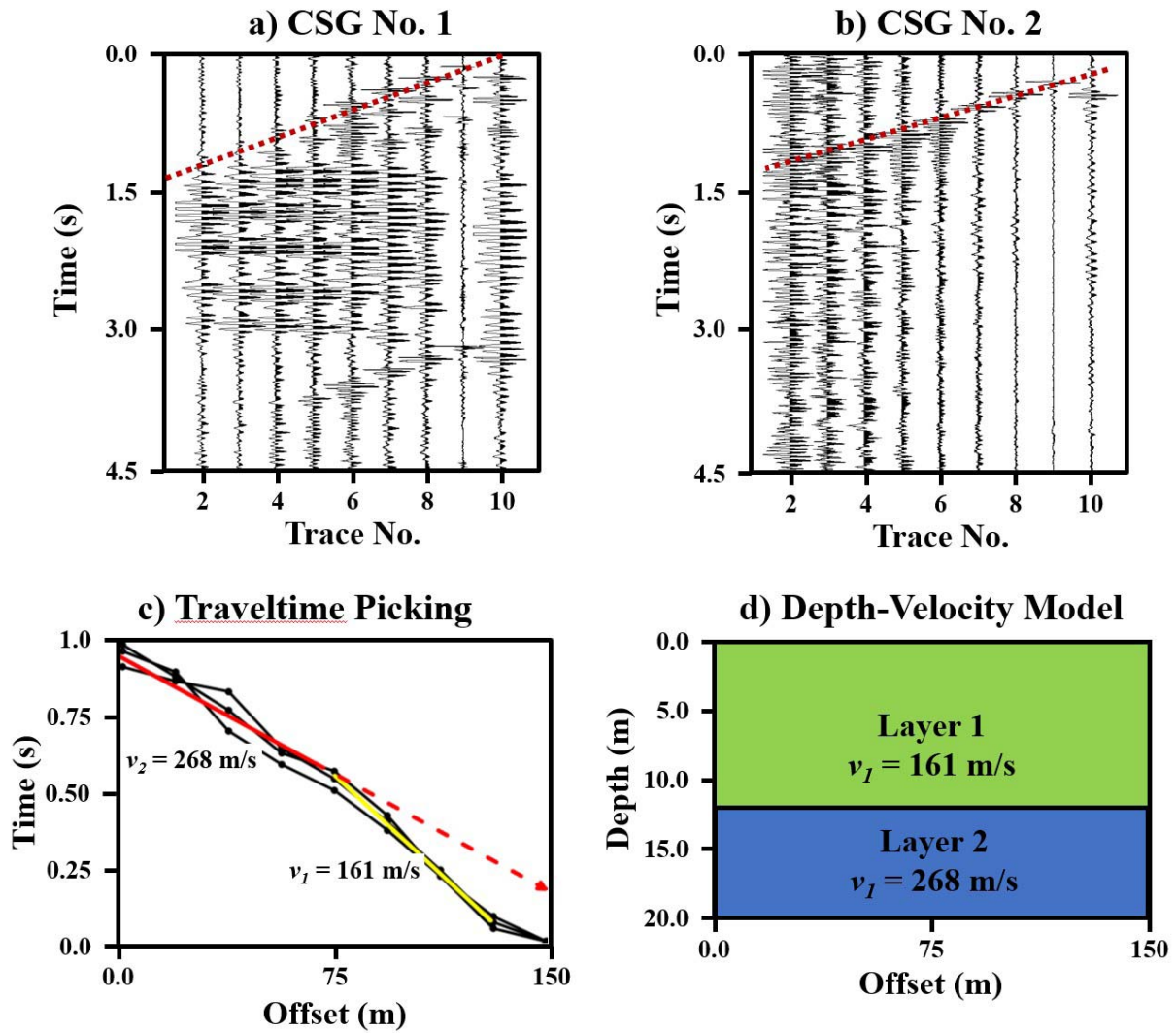

Figure 6: CSGs, traveltimes of first arrivals, and inverted p-velocity model. a) and b) are two CSGs from the recorded active shooting. Both CSGs are excited by hammer blows next to receiver no 10. c) Picked first-arrival traveltimes (FATs) of all three CSGs. d) The generated depth-velocity model using the Intercept Time Method (ITM).

## Discussion

Rocket launches from neighboring spaceports can be recorded and the recorded reciprocal shot gathers can be transformed into a dense set of virtual refraction shot gathers by parsimonious refraction interferometry (PRI). The first-arrival traveltimes from this dense set of refractions can be inverted for the subsurface velocity. For example, Figure 7 depicts a seismic recording line between two neighboring rocket launches which can be used as reciprocal shot gathers. The refraction traveltimes from these seismograms generated by two reciprocal shot gathers can be transformed by parsimonious interferometry to provide a dense set of refraction traveltimes of virtual refraction shot gathers (VRSGs), where a virtual shot[13,14,15,16,17] is at each geophone. For example, the refraction traveltime $T_{AC}$ from the source at A and recorded at C can be added to the traveltime $T_{DB}$ generated at D and recorded at B to give $T_{AC} + T_{DB}$. This sum can then be subtracted from $T_{AD}$ to give the virtual traveltime $T_{BC}=T_{AC}+T_{DB}-T_{AD}$ for a virtual source at B and a receiver at C.

These spaceports must be located within a distance where the signal-to-noise ratio is sufficiently high. On a quiet planetary body such as the Moon or Mars, the distance between neighboring spaceports might be many tens of kilometers. The dense set of VRSGs can be inverted by travel time tomography to provide an image of the subsurface velocity. The surface waves can also be inverted using the PRI approach to give the S-velocity profile[17]. If the spaceports are sufficiently far apart then PRI tomography can provide velocity images as deep as 20 km or more. These will resolve the questions about buried ice deposits below the Martian or lunar surface.

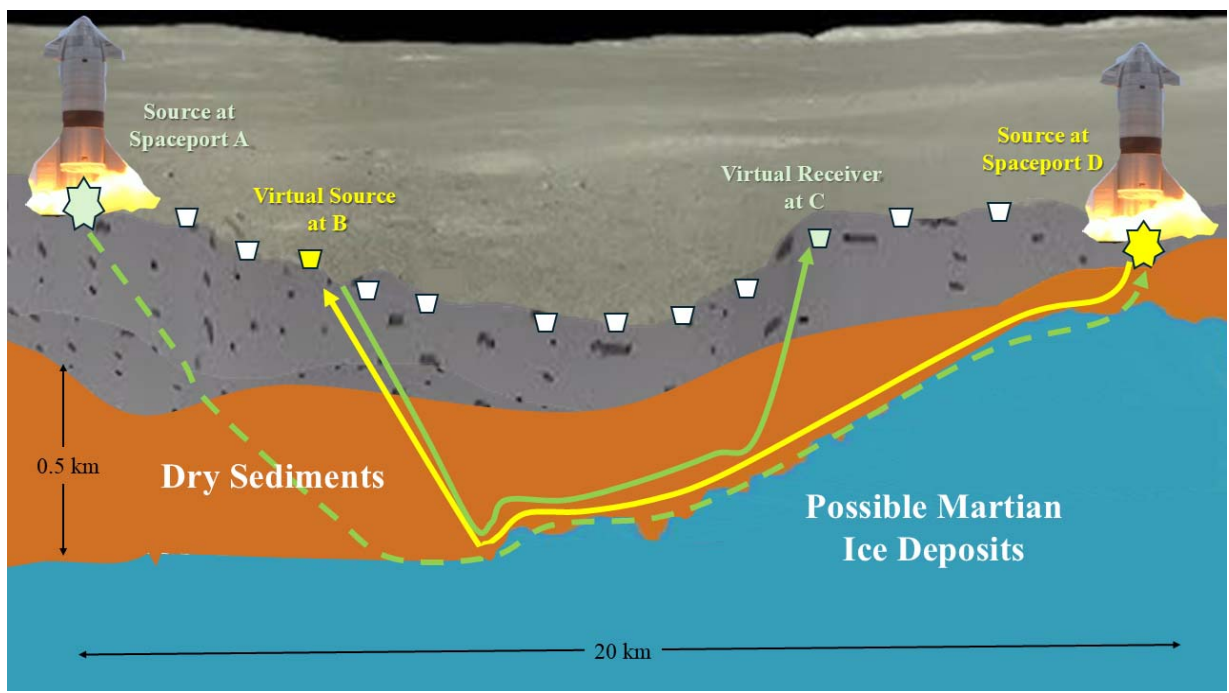

**Figure 7. Rocket sources at spaceports A and D can be used to generate signals recorded by geophones (white quadrilaterals). Using parsimonious refraction interferometry (PRI), virtual common shot gathers of refraction events at the geophones can generated such as the virtual shot at B and receiver at C[13,17]. Illustration adapted from (https://www.esa.int/Science_Exploration/Space_Science/Mars_Express/Buried_water_ice_at_Mars _s_equator) and Yan[11].**

Recent research reveals radar reflections from a geological anomaly between the depths of 0.5-2.7 km deep, which are interpreted as a water-ice layer at that depth[18]. Figure 7 depicts a version of their proposed water-ice stratigraphy on Mars, which is an interpretation of the lithology but not unambiguous proof. We suggest that a reciprocal pair of rocket launches will provide seismic data that can be inverted for the velocity profile that can either support their ice-water hypothesis or refute it. The velocity signature of pure ice water is very distinctive at water = 3.6 − 4.0 km/s[19,20] but can be lowered to 2.4-3.2 km/s if entrained with impurities. For imaging depths of 0.5-3 km, the launch pads do not need to be more than 10 km apart.

# Summary

We tested the feasibility of using a rocket launch, i.e. rocketquake, as a seismic source to image the subsurface velocity and geology of planetary bodies. In our example, the seismic source is the launch of a Falcon 9 rocket at Vandenberg Space Force Base and the seismic recorders consist of 10 passive seismometers. The important findings from this study are the following.

1. Body-wave arrivals at 7 km offset from the launch pad are identified. Their 1.9-2.4 km/s apparent velocities suggest that they dive to a depth greater than 1 km. This means that reciprocal pairs of rocket launches or landings at neighboring spaceports can be used to detect subsurface lithology of interest such as ice-water deposits, tunnels, and hazards. Deploying and hammering the geophones before the launch allows for the recording of P-wave refractions and surface waves that can be inverted for the shallow P- and S-velocity models.

2. The Rayleigh waves excited by the rocket engine can be interferometrically transformed into virtual shot gathers that can be inverted for the subsurface shear velocity to depths greater than 60 m for our data. Inversion of both the vertical- and horizontal-data components give similar shear-wave velocity profiles. This suggests that distributed acoustic sensors, which only record inline horizontal components of data, will provide useful $V_s$ information about the subsurface. The horizontal-component seismograms did not contain visible body-wave arrivals.

3. There will be a growing number of spaceports around the world. As of 2023, there are 20 FAA-licensed spaceports in the USA (https://www.faa.gov/space/spaceports by state). Their rocket launches can be used as a seismic source where geophone lines can be deployed along different radial directions. Such data can be used to detect, for example, the presence of hidden faults in earthquake-prone areas. Rocket launches from neighboring spaceports can combine seismic recordings from different launches and use parsimonious refraction interferometry to provide a more comprehensive image of velocity anomalies. Lunar and Martian buggies should be designed to facilitate the transportation of recording devices.

4. It took decades for earthquake seismologists to partially understand the physics of earthquake sources. In comparison, the physics of rocket sources and the dynamics of their exhaust plumes and acoustic radiation patterns are well understood and tested in the lab. This allows seismologists to use rocket parameters such as acoustic radiation and plume patterns to rapidly explore the potential of using rocket-generated seismic data to image the subsurface geology of planets and their moons[21]. The 1st stage of a Falcon 9 rocket is a free seismic lunch[8] which generates 8277 kN of force in a vacuum (https://en.wikipedia.org/wiki/Falcon_9. This compares to the expensive transportation and use of the largest vibroseis truck which weighs 45 tons and only generates 401 kN of force (https://en.wikipedia.org/wiki/Seismic_vibrator).

# References

1. Mathews, L., Gee, K., Hart, G., Rasband, R., Novakovich, D., Irarrazabal, F., Vaughn, A. & Nelson, P.

Comparative analysis of noise from three Falcon 9 launches. *J. Acoust. Soc. Korea* **39**, 322–332 (2020).
2. Durrant, J., Anderson, M., Bassett, M., Gee, K. & Hart, G. Overview and spectral analysis of the Falcon-9 SARah-1 launch and reentry sonic boom. *184th Meeting of the Acoustical Society of America*, 51, 1–11 (2023).
3. Jardetzky, W. & Press, F. Rayleigh-wave coupling to atmospheric compression waves. *Bull. Seismol. Soc. Am.* **42**, 135–144 (1952).
4. Yilmaz, O. *Seismic Data Analysis* (SEG Press, 2001).
5. Sorlien, C., Kamerling, M. J. & Mayerson, D. Block rotation and termination of the Hosgri strike-slip fault, California, from three-dimensional map restoration. *Geology* **27**, 1039–1042 (1999).
6. Sorlien, C. & Luyendyk, B. Miocene extension and post-Miocene transpression offshore of south-central California. In *Evolution of Sedimentary Basins, Onshore Oil and Gas Investigations - Santa Maria Province* (ed. Keller, M. A.) 38–50 (U.S. Geol. Surv. Bull. 1995-Y, 1999).
7. Lognonne, P., Karakostas, F., Rolland, L. & Nishikawa, Y. Modeling of atmospheric-coupled Rayleigh waves on planets with atmosphere: From Earth observation to Mars and Venus perspectives. *J. Acoust. Soc. Am.* **140**, 1447–1468 (2016).
8. Schuster, G. Free seismic lunches on the Moon with Falcon rocket sources. *Expanded Abstracts for the Fourth International Meeting for Applied Geoscience & Energy* (Houston, Texas, 2024).
9. Novoselov, N., Fuchs, F. & Bokelmann, G. Acoustic-to-seismic ground coupling: coupling efficiency and inferring near-surface properties. *Geophys. J. Int.* **223**, 144–160 (2020).
10. Bensen, G., Ritzwoller, M., Barmin, M., Levshin, A., Lin, F., Moschetti, M., Shapiro, N. & Yang, Y. Processing seismic ambient noise data to obtain reliable broad-band surface wave dispersion measurements. *Geophys. J. Int.* **169**, 1239–1260 (2007).
11. Yan, Y., Chen, X., Huai, N. & Guan, J. Modern inversion workflow of the multimodal surface wave dispersion curves: staging strategy and pattern search with embedded Kuhn-Munkres algorithm. *Geophys. J. Int.* **231**, 47–71 (2022).
12. Hanafy, S., AlTheyab, A. & Schuster, G. T. Controlled noise seismology. *SEG Tech. Program Expanded Abstracts*, 5906063 (2015).
13. Hanafy, S. & Schuster, G. T. Parsimonious refraction interferometry and tomography. *Geophys. J. Int.* (2017).
14. Li, F., Hanafy, S. & Schuster, G. T. Parsimonious wave-equation traveltime inversion for refraction waves. *Geophys. Prospect.* 1–10 (2017).
15. Li, J., Hanafy, S. & Schuster, G. Parsimonious surface wave interferometry. *Geophys. J. Int.* **212**, 1536–1545 (2017).
16. Li, J., Hanafy, S. & Schuster, G. Wave-equation dispersion inversion of guided P waves in a waveguide of arbitrary geometry. *J. Geophys. Res.* (2018).
17. Hanafy, S. M., Hoteit, H., Li, J. & Schuster, G. Near-surface real-time seismic imaging using parsimonious interferometry. *Sci. Rep.* **11**, 7194 (2021).
18. Watters, T., Campbell, B., Leuschen, C., Morgan, G., Cicchetti, A., Orosei, R. & Plaut, J. Evidence of ice-rich layered deposits in the Medusae Fossae formation of Mars. *J. Geophys. Res.* **51**, 1–12 (2024).
19. LeBlanc, A., Fortier, R., Allard, M., Cosma, C. & Buteau, S. Seismic cone penetration test and seismic tomography in permafrost. *Can. Geotech. J.* **41**, 796–813 (2011).
20. Baker, G., Strasser, J., Evenson, E., Lawson, D., Pyke, K. & Bigl, R. Near-surface seismic reflection profiling of the Matanuska Glacier, Alaska. *Geophysics* **68**, 147–156 (2003).
21. Schuster, G., Acoustic Green's function for an ascending rocket (in preparation): arXive (2025).

# Appendix 1

The spectra for the Z-component recordings are in Appendix's Figure 1, where there is useful information from about 0.2 Hz to more than 50 Hz. Appendix's Figure 2 depicts the seismograms recorded at the

VSFB prior to the launch.

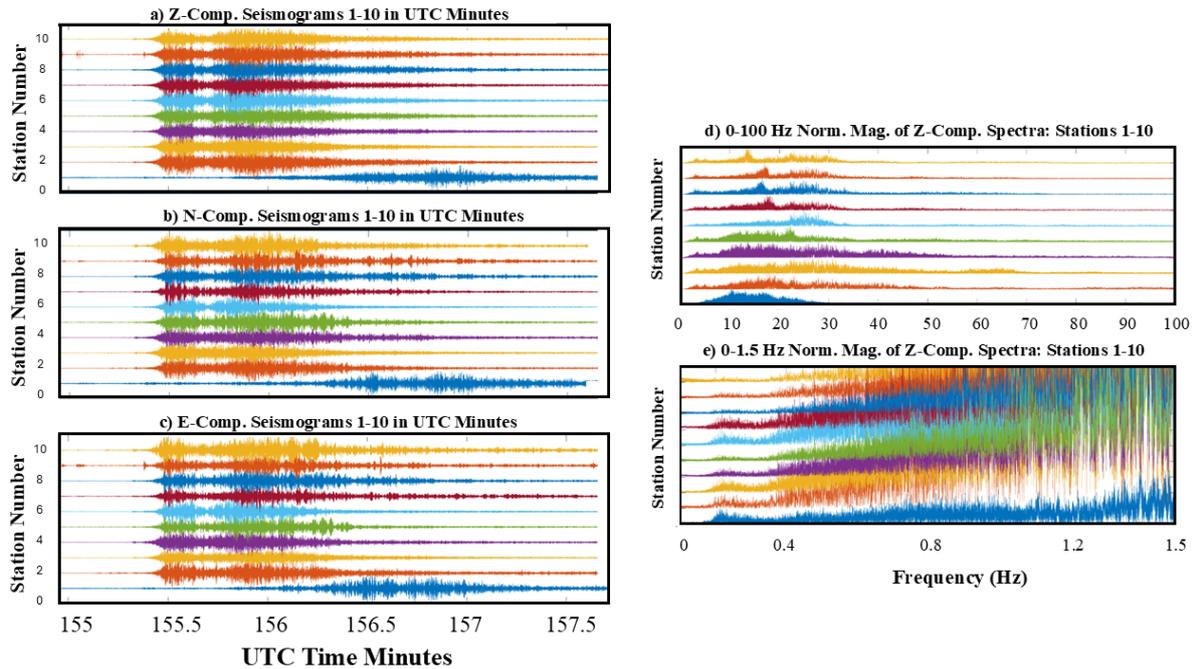

**Appendix 1.** Three-component seismograms and their spectra. (a-c) Three-component recordings at stations 1-10, where the rocket launch started at 155 minutes UTC time (7:35 PM California time). Station 1 is in the town of Lompoc, which is about 15 km from Launch Pad 4. The Z-component magnitude spectra are shown in d)-e).

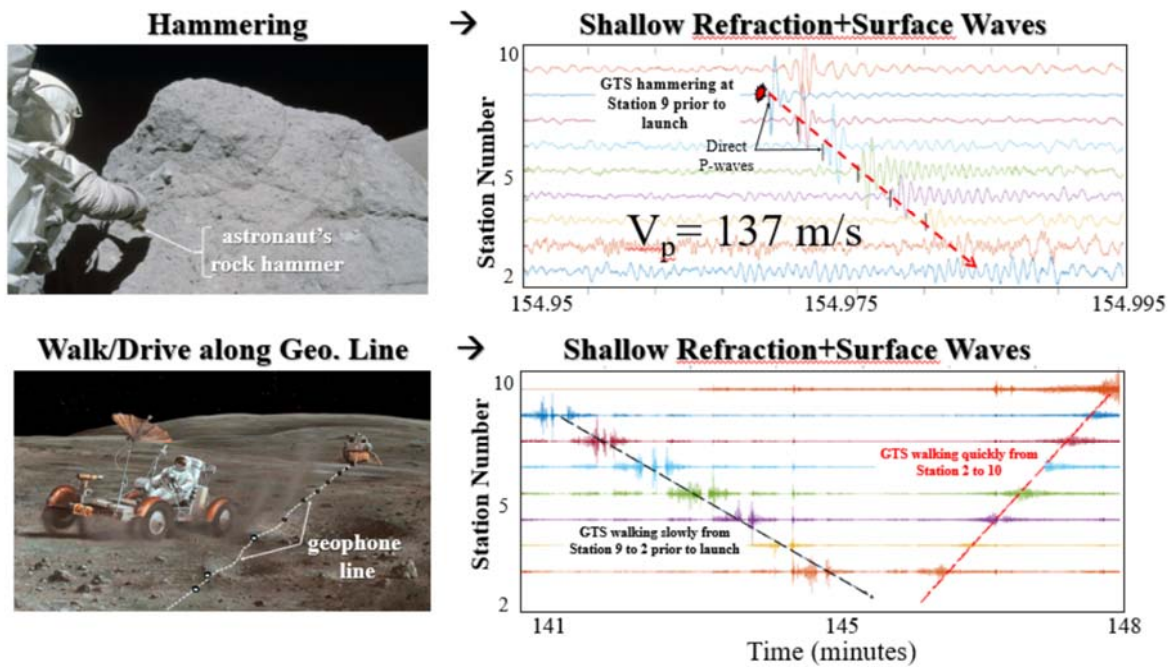

**Appendix Figure 2.** Seismograms generated by (top) hammering at station 9 with a rock hammer and b) walking from station 10 to and from station 2. Top seismograms can be inverted for the shallow VP and VS structures and faults. The bottom seismograms can be interferometrically transformed into virtual CSGs and the surface waves inverted for the deeper Vs distribution.